# Fermilab experience of post-annealing losses in SRF niobium cavities due to furnace contamination and the ways to its mitigation: a pathway to processing simplification and quality factor improvement

A. Grassellino, A. Romanenko, A. Crawford, O. Melnychuk, A. Rowe, M. Wong, C. Cooper, D. Sergatskov, D. Bice, Y. Trenikhina, L.D. Cooley, C. Ginsburg, R. D. Kephart

*Abstract*
We investigate the effect of high temperature treatments followed by only high-pressure water rinse (HPR) of superconducting radio frequency (SRF) niobium cavities. The objective is to have a cost effective alternative to the typical cavity processing sequence, by eliminating the material removal step post furnace treatment while consistently preserving or improving the RF performance of fine grain and large grain cavities. The studies have been conducted in the temperature range 800-1000°C for different conditions of the starting substrate: large grain and fine grain, electro-polished (EP) and centrifugal barrel polished (CBP) to 'mirror finish'. An interesting effect of the grain size on the performances is found. Cavity results and samples characterization show that furnace contaminants cause poor cavity performance, and a practical solution is found to prevent surface contamination. Extraordinary values of residual resistances ~ 1 nΩ and below are then consistently achieved for the contamination-free cavities. These results lead to a more cost-effective processing and improved RF performance, and, in conjunction with the new CBP results, open a potential pathway to acid-free processing.

*Introduction*

Many modern and proposed future particle accelerators rely on superconducting radio frequency cavities made of bulk niobium as primary particle accelerating structures. The state-of-the-art processing stream for SRF niobium cavities includes bulk material removal via electro-polishing (EP), buffered chemical polishing (BCP) or centrifugal barrel polishing. All these techniques can introduce significant amount of hydrogen in the cavity, therefore a high temperature degassing treatment, in the range 600-1000°C, is necessary to prevent poor cavity operational performance due to formation of large hydrides precipitates Nb-H, phenomenon known as 'Q-disease' [1]. The heat treatments are then followed by a 20-40 μm amount of material removal again with EP or BCP, typically done to remove surface 'contaminants' which might be introduced during the furnace treatment. In an effort to obtain a more cost-effective processing of niobium SRF cavities, we initiated a series of experiments with the goal of gaining an understanding of the nature of the surface contamination and of eliminating this post-degassing material removal step. A further motivation for pursuing this was trying to obtain a niobium cavity surface with reduced amount of hydrogen, and to investigate the effect on its RF performance. Typically the material removal via chemical treatments may reintroduce hydrogen into niobium as found out by previous studies [2]. Some heat treatment studies with no subsequent material removal on few large grain cavities and one fine grain cavity [3, 4] showed increased cavity quality factor ($Q_0$) values, further motivating these investigations, as it was also suggested in [5, 6]. One interesting point is that different results have been found for fine grain cavities post-heat treatment with no subsequent material removal, showing often very poor performance [7, 8, 9]. To date there has not been a detailed study of the possible roots of these poor results, and of a potential solution to achieve consistently good performance on all cavities annealed with no subsequent chemistry (fine grain, large grain, differently processed substrates), which strongly motivated this work at FNAL. Finally, there has been an extensive effort at FNAL in producing an acid-free material removal process, via centrifugal barrel polishing (CBP), material removal technique that allows obtaining 'mirror smooth' surfaces [10]. However, CBP loads the cavity with large amount of hydrogen and a final degassing step followed by light chemistry is always needed post CBP. This further motivated our studies, since the elimination of the post-furnace material removal is essential to make CBP a completely chemistry-free procedure and to preserve the mirror smooth finish.

*Experimental tools*
A picture of the T-M Vacuum Furnace used for these studies is shown in Figure 1. The chamber can accommodate two single cell cavities or a 1.3 GHz nine-cell cavity. Two cryopumps are attached to the chamber and provide a total pumping speed of 9600 L/sec air and 24,000 L/sec hydrogen. A dry Roots pump provides rough pumping capability. Throughout the furnace operation, RGA measurements of the partial pressures are recorded. A full spectrum of 1-100 amu is recorded every minute. Cold cathode gauges provide total pressure readings. The maximum allowed operating temperature of the furnace is 1000°C. The heating elements are 2-inch wide molybdenum strips which surround the hot zone, which is a volume of 12" x 12" x 60". Five layers of molybdenum make up the thermal shields. Chilled water kept at 70°F keeps the outer surface of the dual-walled vacuum chamber cool to the touch during operation.

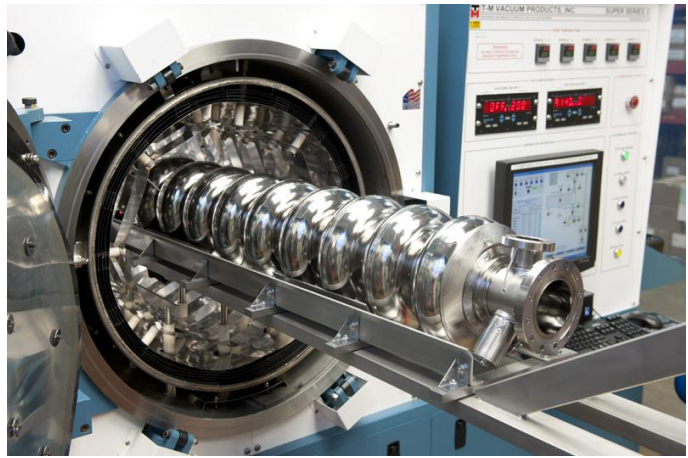

Figure 1. Picture of the T-M furnace used for the heat treatment studies of niobium cavities, with a 9-cell 1.3 GHz cavity being loaded in the chamber.

All the cavities used in these studies are 1.3 GHz TESLA shape cavities [11], with Nb-Ti alloy flanges. The serial number and some parameters of these cavities are summarized in Table 1. The typical electro-polishing removal is done using a standard solution of $H_2SO_4$:HF 9:1, while details of the material removal via CBP can be found in [10].

The cavity performance is characterized via a standard power measurement of the quality factor Q versus RF field. Most of the tests presented in the paper at the bath temperature of 2 K (unless specified), and we use fixed input couplers with $Q_{ext1}$ values ranging from $6\times10^9$ to $1\times10^{10}$ producing an error in the Q measurements ~10-20%, depending on the case by case mismatch with the cavity Q. No field emission was present in any of the presented cavity tests unless otherwise noted. Some of the cavity tests were performed with a custom built temperature mapping system attached to the outside wall (shown in Figure 2), similar to that described in [12].

Table 1: List of cavities used for the experiments

| CAVITY ID | Type | Substrate treatment |
| --- | --- | --- |
| TE1AES016 | Large grain | EP |
| TE1AES003 | Fine grain | BCP |
| TE1AES005 | Fine grain | EP |
| TE1PAV005 | Fine grain | EP |
| TE1PAV006 | Fine grain | EP |
| TE1AES008 | Fine grain | CBP |
| 1DE20 | Large grain | CBP |
| TE1ACC001 | Fine grain | EP |
| TE1AES013 | Fine grain | EP |
| PIPPS003 | Fine grain | EP |

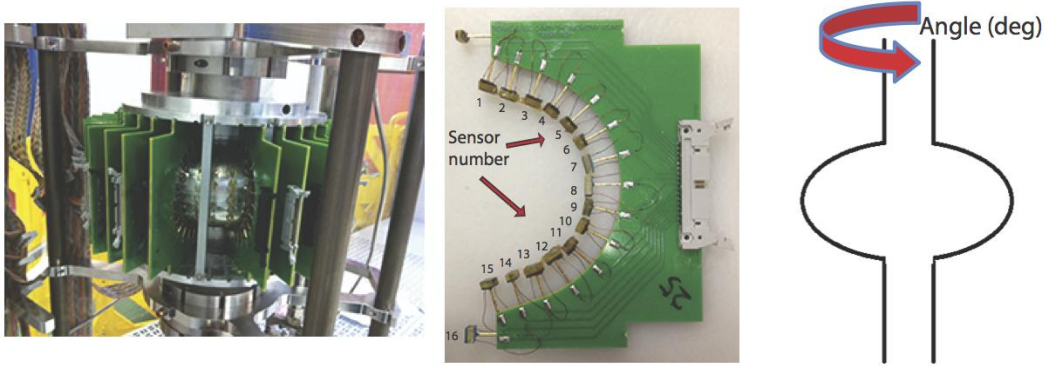

Figure 2. Left: Temperature mapping attached to the outside cavity walls. Center: individual board with sensor numbering. Right: schematic showing how angle is measured around the rotational cavity axis, 36 boards are uniformly spaced with 10° separation.

*Cavity measurements*
We started the studies by applying no chemistry after 800°C 3 hours vacuum furnace treatment on a large grain (~10 cm) EP cavity TE1AES016. The large grain cavity results are shown in Figure 3. After the 800°C treatment followed by ultrasonic cleaning and HPR only, the $Q_0$ of the cavity increased significantly at all field levels compared to the baseline EP Q curve. The increase in low field Q stem from an improvement in residual resistance, which decreased from ~ 4 nΩ at baseline to ~ 1.5 nΩ, as calculated via Q versus temperature measurement at 5 MV/m.

Subsequent 120°C 48 hours baking decreased the $Q_0$, due to the increase in the residual resistance, while BCS resistance remained unchanged. A hydrofluoric acid (HF) rinse for 5 minutes applied after the 120°C treatment restored the $Q_0$ to practically the same level as before, which is consistent with the other studies reported in [13]. Temperature mapping of the RF losses was used after the cavity received a 120°C bake plus an HF rinse, and a temperature map at $E_{acc}$ = 22 M/m is shown in Figure 4. The losses causing the medium field Q-slope are concentrated mostly in the region of a high magnetic field around cavity equator in line with findings in [13] as well. No anomalous losses at any field levels were found post the high temperature bake followed by HPR only.

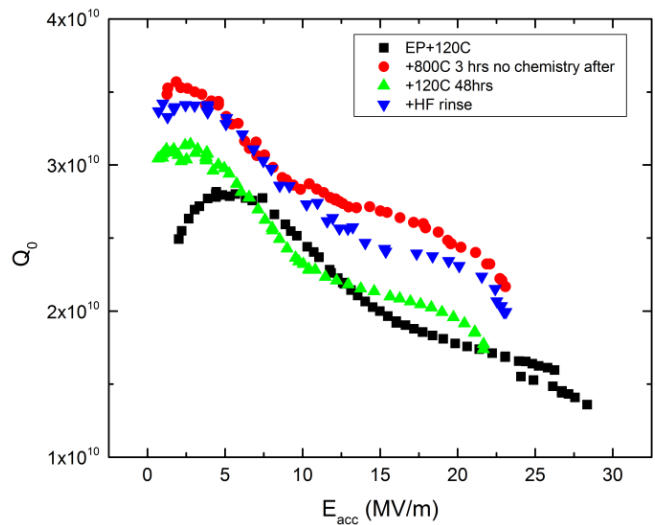

Figure 3. $Q_0$ ($E_{acc}$) curves at T = 2 K for the large grain cavity (grain size ~10 cm). In sequence: black curve for baseline EP, red after 800°C 3 hours, green after 120°C bake, and blue after one HF rinse. Notice the linear scale.

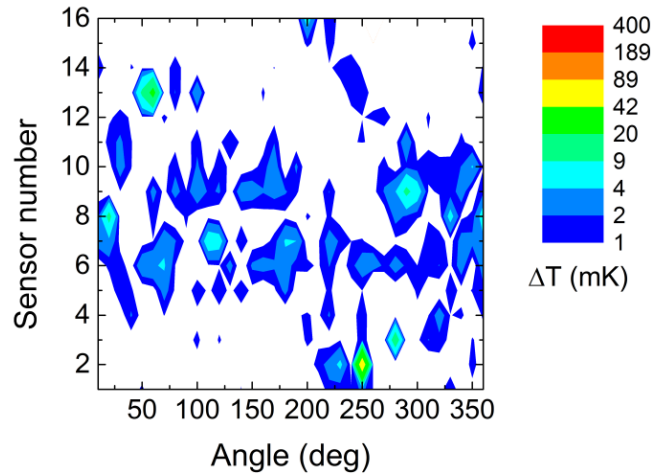

Figure 4. Contour plot of the temperature map at $E_{acc}$ = 22 MV/m for the large grain EP cavity (blue curve in Fig. 1). Color scale is in Kelvin.

We continued the studies by applying no material removal post 800°C treatment to several fine grain cavities with different starting substrate treatments, as described in Table 1. Contrary to the successful large grain result, all fine grain cavities exhibited poor performance after the 800°C furnace treatment followed by no material removal. A severe 'Q-disease-like' behaviour appeared in the $Q_0(E_{acc})$ curves at T = 2 K as shown in Figure 5, even though all the cavities were rapidly cooled down (maximum 10 minutes through the region 175K-80K), and no x-rays were detected. Several Q versus temperature measurements showed that these poor performances were dominated by large residual resistances, as shown for example by the Q vs E curves at 2K and 1.6K for the two cavities TE1PAV006 and TE1AES008 in Figure 6. Due to the shape of the Q-curves resembling the classical hydrogen Q-disease shape, it was at first thought that large amount of residual hydrogen were left in the cavity despite the degassing cycle, as it was suggested in previous work [14]. A 100 K hold of one the fine grain cavities exhibiting poor performance (TE1PAV005) was therefore performed, for about 7 hours. This is a typical test used for revealing the presence of large amounts of hydrogen, since the Nb-H precipitates form in the range ~ 80-175K and grow in larger islands with time spent at this temperature, if hydrogen supply from the bulk is present. The result was that the Q curve post 100 K hold remained unchanged.

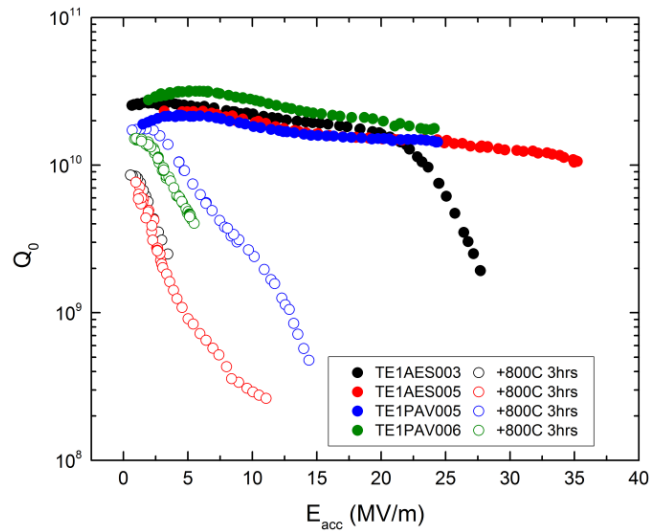

Figure 5. Q-disease-like performance in fine grain cavities after 800°C 3 hours furnace treatment with no material removal afterwards. Solid points are before the high temperature bake and open ones of the same color are after the bake for the same cavity. All curves shown are taken at bath T of 2K.

In order to investigate the origin of the 'Q-disease like' behavior, we performed some material removal studies, to understand to what depth the losses were localized. We first tried 10 HF rinse cycles on the cavity TE1AES005. Such treatment removes of the order of 20 nanometers of the niobium surface. The cavity did not recover and the Q-disease was still present with some slight change as shown in Fig. 7 a. Temperature maps obtained for this RF test have shown heating over the whole cavity surface as shown in Fig. 7 b.

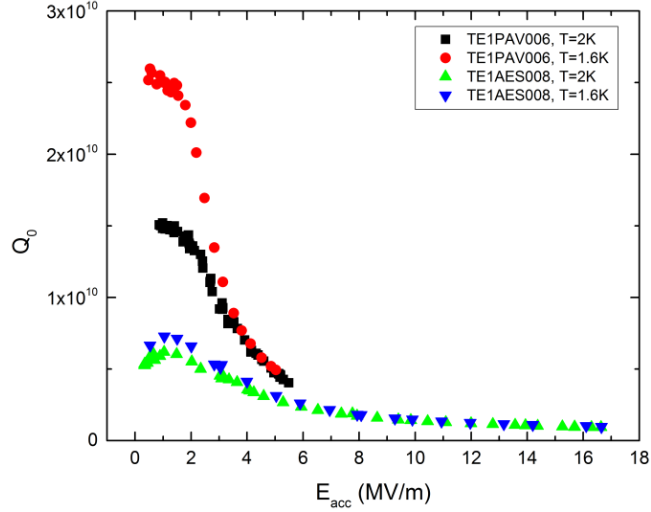

Figure 6. Q vs E curves at two different temperatures 2K and 1.6K, showing that large residual resistances dominate the strongly field dependent surface resistance of the fine grain cavities heat treated at 800°C for 3 hours followed by no material removal.

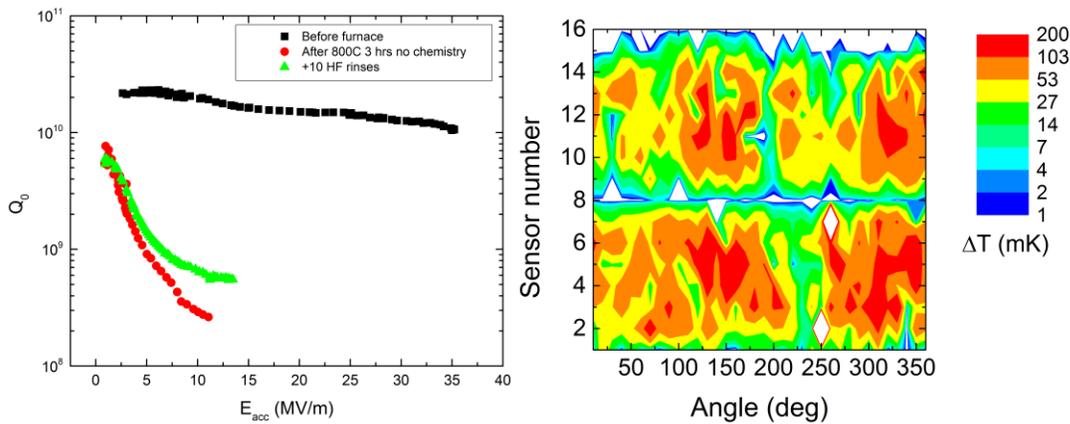

Figure 7. a) Non recovery of the Q-disease for cavity TE1AES005 after 10 HF rinses (2K curves); b) Strong heating of the most of the cavity surface into the Q-disease. Color scale is in Kelvin.

We then performed 20 microns electro-polishing removal on several 'Q-diseased' fine grain cavities, and the performance went back to normal (in some cases quench field was lowered), as shown for example in fig. 8 for the cavity TE1AES003. This explained why this effect was not noticed before, when the typical processing sequence always involved 20 or more microns of material removal post furnace treatment. We later on performed a test on an electro-polished fine grain cavity, which was baked at 800°C for 3 hours and had subsequent 5 microns EP removal. The cavity had a regular Q at 2K ~ 2.5e10 and was limited at 35 MV/m by HFQS. So 5 microns material removal post furnace treatment was found to be enough to restore regular performance.

We then investigated if those losses might have been an effect of backfilling the furnace at room temperature with the argon gas. So we tried room temperature venting of the chamber with different gases: first we tried nitrogen, most typically used for room temperature venting of high temperature furnaces. Room temperature nitrogen venting was first tried on a fine grain cavity that was baked at 1000°C for 36 hours, as shown in fig. 8. In the sequence we see that the cavity TE1AES003 degrades significantly after 800°C 3 hours, venting at room temperature with argon. Then, performances recover after 20 microns removal via EP, except that the quench field is lowered. After the cavity was baked at 1000°C for 36 hours, with venting with nitrogen at room temperature, the Q curve has again a 'standard' shape, although the quench field is lowered even further. This result seemed to show a beneficial effect either of the longer baking cycle at higher temperature, or of the venting with nitrogen instead of argon, or both.

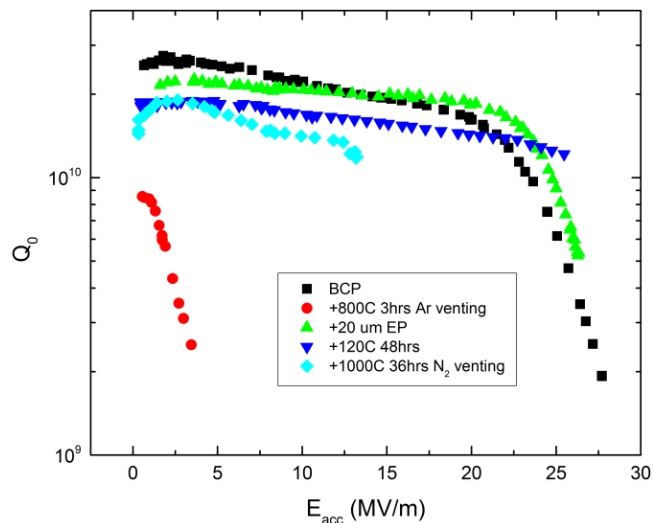

Figure 8. Sequence of furnace and chemical treatments on the same fine grain 1.3 GHz single cell cavity TE1AES003, 2 K Q curves.

To decouple the effect of the different baking temperature and length from the room temperature venting, another single cell cavity was baked at 800°C for 3 hours and vented with nitrogen at room T. The cavity is TE1AES008, which prior to the bake, had received bulk removal via CBP and had been tumbled to mirror finish, without receiving any EP or BCP. Results are shown in fig. 9, indicating that medium field losses were 'milder' compared to the argon vented cavities. The cavity did not quench, it was limited by available power. So there appeared to be a beneficial effect of nitrogen or a detrimental effect of argon venting at room temperature.

We then continued the room temperature venting studies, this time using dry air. The cavity is TE1PAV005, which was previously baked also at 800°C for 3 hours but vented with argon. A comparison of the results is shown again in Fig. 9, showing that the medium field losses were significantly improved, and the cavity could reach 20 MV/m. Even though performances were still poor and 'Q-disease-like', there was a clear improvement from the dry air chamber backfilling compared to argon. Interestingly, the effect of dry air and nitrogen venting on medium field losses is the same, compared to argon.

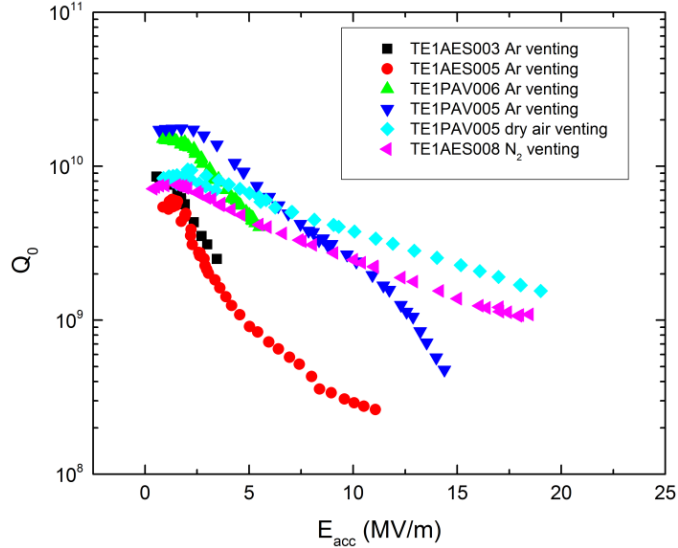

Figure 9. Study of the effect of backfilling the furnace below 50°C (post 800°C bake) with different gases: argon, nitrogen and dry air, Q curves at 2K.

*Heat treated CBP cavities*

The mediocre performance of the tumbled cavity TE1AES008, which had received no material removal via chemistry, seemed at this point to possibly be due mainly to the furnace contamination, and not by residual embedded media left over by the centrifugal barrel polishing. Following our finding of a grain size effect on losses, the large grain 1.3 GHz single cell cavity 1DE20 was then prepared identically to TE1AES008: it was tumbled to mirror finish, degassed at 800°C for 3 hours (room temperature venting with nitrogen), high pressure rinsed and tested. Results are shown in fig. 10 a. The cavity had a good low field Q of ~ 2e10 and a low field residual resistance extracted via Q vs temperature measurement performed at 5 MV/m of ~ 3 nΩ. Medium field losses were stronger than standardly treated cavities, but milder than all the argon vented cavities.

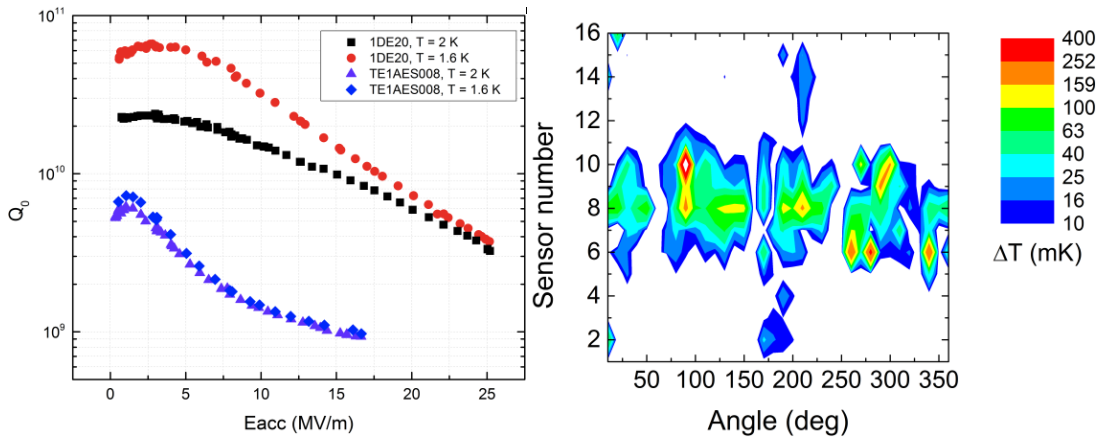

Figure 10. a) Q curves at 2K and 1.6K for the large grain cavity 1DE20, tumbled to mirror finish, baked at 800$^0$C for 3 hours and for comparison of TE1AES008, same treatment but fine grain; b) T-map of 1DE20 at medium field, showing heating at the equator, typical signature of hydrogen Q-disease.

A comparison with the fine grain cavity TE1AES008 which was treated identically is shown in fig. 10 a. Performance of the large grain cavity were by far superior to the fine grain, and low field residual resistance was 3 nΩ for the LG versus 40 nΩ for the FG cavity! This confirmed again the grain size effect on the post-furnace treatment losses, and that indeed the poor performance of the tumbled fine grain cavity was dictated by the post-heat treatment losses and not by tumbling embedded media.

The 1DE20 results represent a first and extremely encouraging data point towards the possibility of acid-free processing. Past attempts on minimization of material removal post-CBP showed that some amount of BCP/EP is necessary to avoid extremely bad performance [6]. Even though results are encouraging, it's important to point out that the losses in the medium field in 1DE20 were stronger than in typical electro-polished cavities, for example compared to TE1AES016, and could be caused by some residual media embedded by CBP. We then studied the 1DE20 cavity losses via T-map (fig. 10 b) and found that all the losses were concentrated at the equator – typical signature of hydrogen Q-disease [1]. CBP typically loads large amount of hydrogen in the cavity, so it could be possible that 3 hours at 800°C were not sufficient to remove all the hydrogen from the cavity. The RGA data of the degassing cycle of the tumbled large grain is shown in fig. 11 a. The cavity was then subsequently heat treated at 800°C for 6 hours and re-tested, but performance degraded, as shown in fig. 11 b. Further investigations on the cause of these stronger medium field losses in CBP surfaces (followed by no EP/BCP) are currently under investigation and will be subject of future publications.

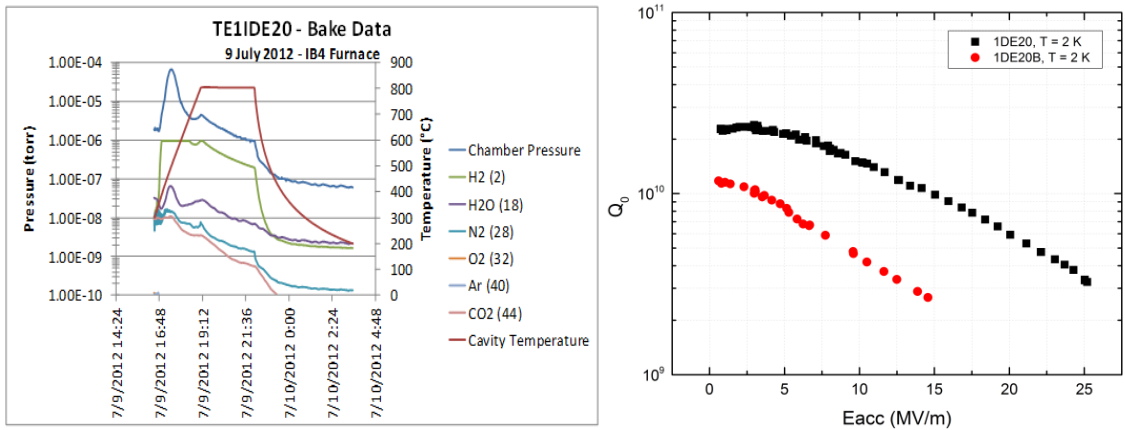

Figure 11. a) RGA data for the 3 hours degassing cycle of 1DE20; b) performance comparison of the large grain tumbled cavity 1DE20 post 3 hours at 800$^0$C and subsequent longer degassing cycle 6 hours at 800$^0$C.

*Samples studies*
Following the series of tests described so far, it emerged that some form of contamination was located in the first 5 microns or less of the cavity surface, causing poor cavity quality factors and large residual resistances. A beneficial effect of nitrogen and dry air venting was found over argon venting of the furnace at room temperature. A grain size effect was found, and also baking for 36 hours at 1000°C produced better results. We started then some samples studies to gain a better understanding of the possible source of these findings. Several fine grain niobium samples, which had received bulk removal via either electro-polishing or buffered chemical polishing, were baked at 800°C for 3 hours and some at 1000°C for 10 hours. The 3-mm thick and ~ 1 cm squared samples are placed on a Nb tray with one side facing up towards the heaters and one laying on the tray. The samples are studied first via laser confocal microscopy at room temperature. Interesting features were found consistently on the samples, which are shown in the laser confocal microscope images in Fig. 12 and 13. All the samples post 800°C bake, were contaminated by precipitates at the grain boundaries and also by some intra-grain precipitates for certain grains only (fig.12). For the samples that were baked at 1000°C, the grain boundaries contamination appearance changed from granular precipitates to a less discontinuous structure (fig.13). Interestingly, the grain boundary contamination was found always only on one side of the samples, while the other remained clean (see for comparison bottom right picture in fig.12).

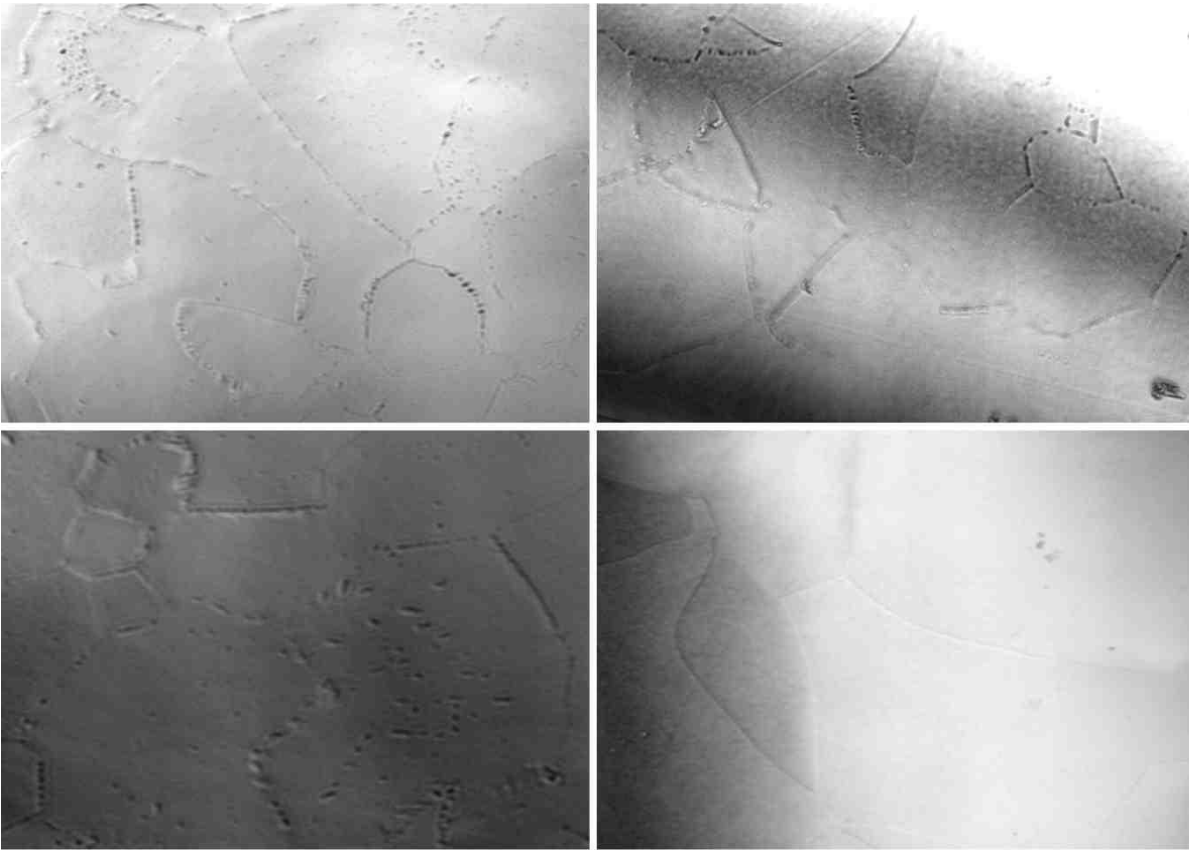

Figure 12. Contaminated grain boundaries of fine grain Nb electro-polished samples post furnace treatment at 800°C for 3 hours. Some intra-grain precipitation is also found. The bottom right picture shows for comparison the second side of the sample (the one facing not facing the heaters during the bake), which shows clean grain boundaries and no intra grain precipitates.

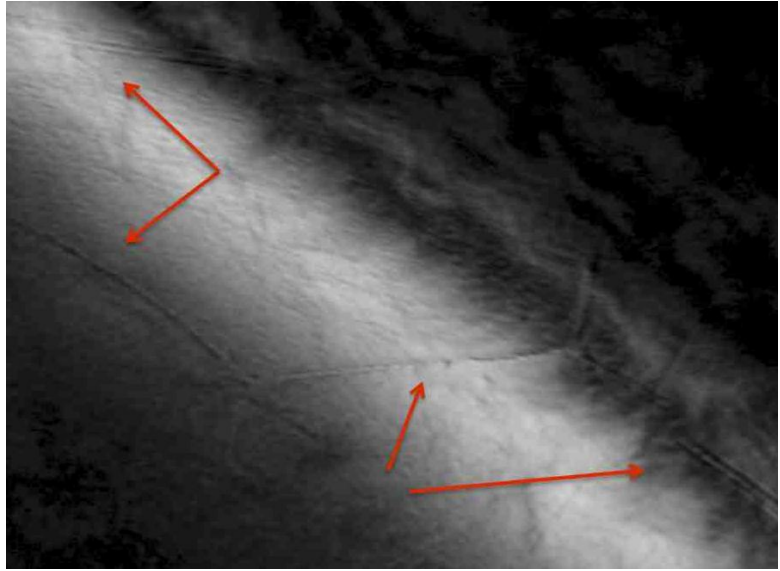

Figure 13. Contaminated grain boundaries of fine grain Nb electro-polished samples post furnace treatment at $1000^0$C for 10 hours. Precipitates now form a less discontinuous structure along the grain boundaries.

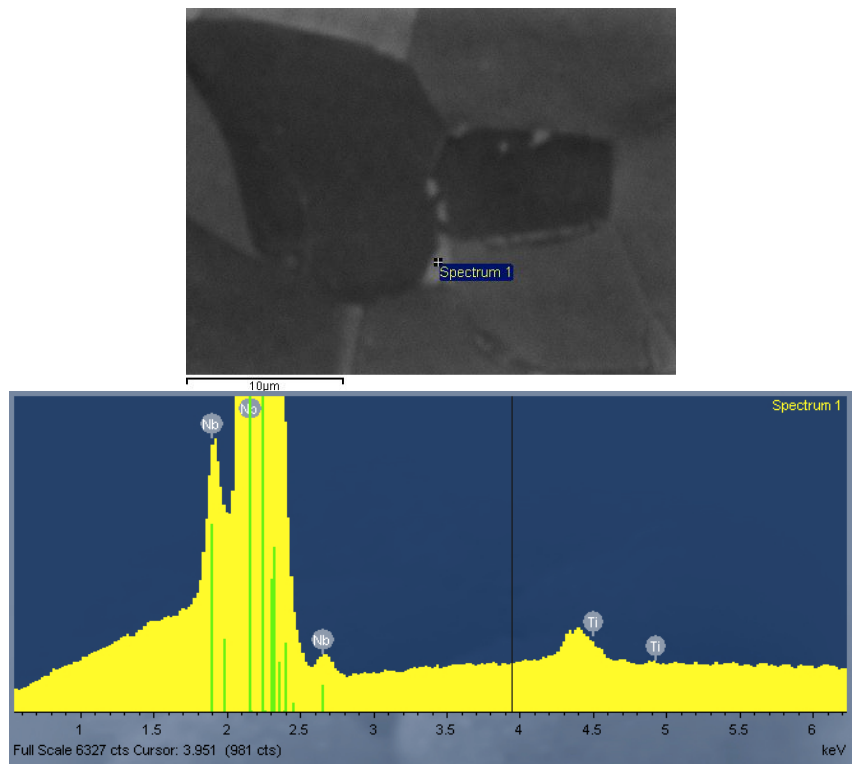

Figure 14. EDX results for a polycrystalline EP sample treated in the furnace at $800^0$C, showing a diffracting peak corresponding to titanium. The top SEM picture shows the area (marked 'spectrum 1') from which the EDX spectrum was collected, which corresponds to precipitates at grain boundaries. The area sampled with EDX is a circle of radius ~ 1 micron.

The heat treated samples were then analysed via laser confocal microscopy at cryogenic temperatures. A cryo-stage allows cooling the sample to 4K and taking images in real time [15]. After holding the samples at 100 K for some time, no formation of hydrides was observed, indicating that no large quantity of hydrogen is present in the samples. Samples were also studied via EDX and results showed presence of a diffracting peak that coincides with the primary peak for titanium (see fig. 14). Some XPS studies (fig.15) on a fine grain 800°C heat-treated sample also confirmed the presence of titanium [16]. This finding is not too surprising, since the only other element that is baked in the furnace is titanium, in the Nb-Ti alloy that makes the cavity flanges. The furnace never exceeded 1000°C, but it could be a high enough temperature to evaporate titanium from the flanges and cause furnace contamination. This finding can provide a potential explanation for most of the cavity results. Q-disease has been typically known as a manifestation of large amount of hydrogen in niobium, which in the temperature range 100 K-160 K forms Nb-H precipitates. Those precipitates are superconducting only via proximity effect [17] and with increasing field cause the classic 'Q-disease' shape. But this behaviour would be the same for any form of weakly superconducting precipitates, which in this case could be α-Ti, or some carbide of niobium and titanium. This can explain also the grain size effect, since precipitates will form mostly at grain boundaries and at lattice irregularities, which fine grain cavities, differently from large grain ones, are rich of.

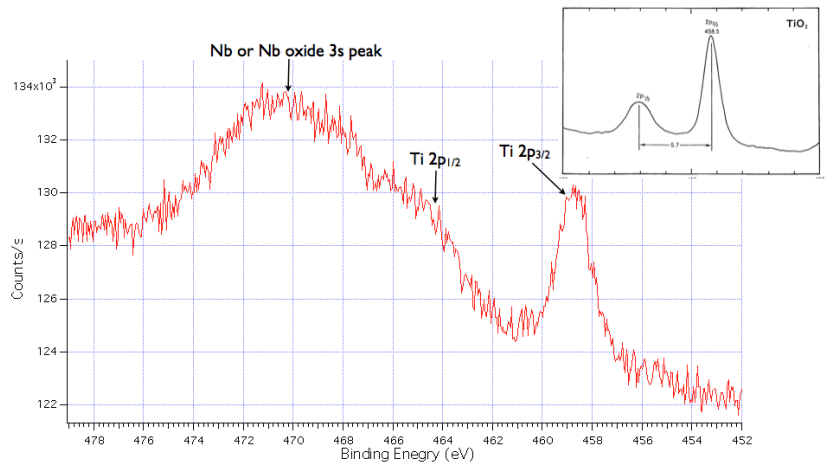

Figure 15. XPS spectrum for a polycrystalline EP sample treated in the furnace at 800°C, showing energy peaks corresponding to titanium dioxide [16].

*Baking with end caps: a practical solution*
Once it was established that there was titanium contamination in the furnace, a solution was sought to prevent titanium in the chamber to land and stick to the cavity internal surface during the bake cycle. We designed a set of niobium beamtube caps, shown mounted on the cavity in fig. 16 a, which wrap around the cavity flanges and exclude any line of sight to the cavity inner surface. Some of the cap fingers that bend around the flanges are designed to leave small opening slits for giving a pathway out for desorbed gases. A similar preventive solution was already described in [18]. Comparative results of a processing sequence on the same fine grain electro-polished cavity TE1ACC001 are shown in fig. 17.

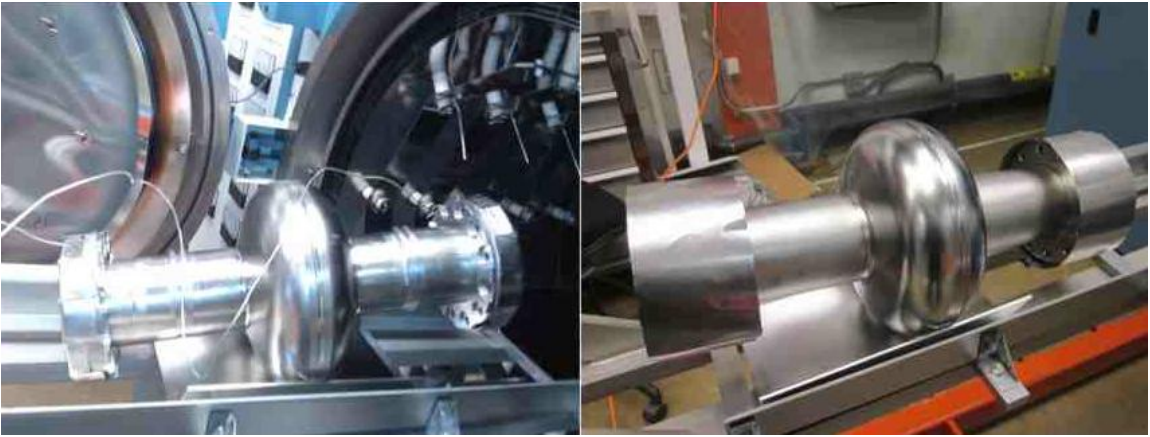

Figure 16. a) Cavity going in the furnace with the Nb beamtube caps; b) beamtube caps plus wrap-around-foil solution which improved performance even further.

The cavity was first baked at 800°C for 3 hours with no caps, resulting in the typical Q-disease performance. Then the cavity received 20 microns removal via electro-polishing, resulting in a 'reset' of the surface, with standard performance, limited by a multipacting induced quench at ~ 20 MV/m. The cavity was then baked again at 800°C for 3 hours, this time with the beamtube caps described above. Performances were clearly and significantly improved, showing a high Q at low field with a moderate slope at medium field, as shown in fig. 17. It is interesting to notice that this cavity TE1ACC001 did not show a clear onset of high field Q-slope post 800°C bake up to the quench field of 32 MV/m, even though it had not received a 120°C bake. We then prepared two more fine grain cavities (PIPPS003 and TE1AES013) with the same technique, but we added an extra protection layer via some Nb foil as shown in fig. 16 b. For TE1AES013 the chamber was backfilled at room temperature with dry air, while for the other cavity with nitrogen. Results of the tests can be found in fig. 18. A summary of the Q and residual resistances values extracted via Q versus temperature measurement at 5 MV/m can be found in table 2. Both cavities baked with caps and foil showed a good Q and mild medium field Q-slope. So the caps solution proved again to be effective, and the additional foil helped eliminating the medium field losses compared to the previous results with TE1ACC001. This indicates that some contamination (in much smaller quantity) was still deposited on the cavity surface through the caps degassing slits. The extra shield of foil helped eliminating it, leading to cavity performance actually even superior to what obtained with standard EP- as shown in fig. 18 for comparison with a standardly treated EP cavity, at 2 K bath temperature. The root of these superior performances can be found in table 2: excellent residual resistances values were achieved in all the fine grain cavities baked with caps and in the large grain cavity even if baked with no caps, post 800°C treatment. All of these low field residual resistance values are significantly below average ones for a standard EP 1.3 GHz cavity, which is typically around 4 nΩ. It is also interesting to notice that the onset of HFQS is higher than typical for an EP cavity in the case of TE1AES013, which was vented with dry air. PIPPS003 showed some multipacting and multipacting induced quenches at 16 MV/m, which caused some Q degradation due to trapped flux, and which quickly processed away. It is also very interesting that no field emission was observed in any of these tests up to the highest gradients achieved. This is an important point, since field emission was considered one of the potential limitations for high temperature treatments without subsequent chemistry.

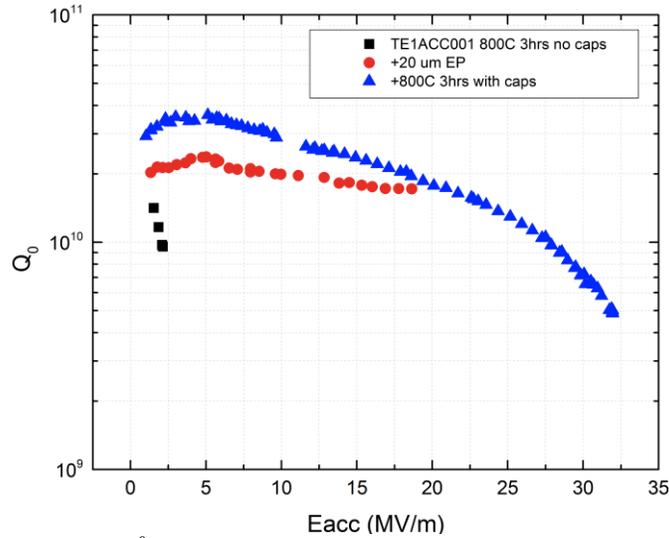

Figure 17. Comparison of post $800^0$C performance with no caps, after 20 micron EP and finally after $800^0$C with baking caps (2 K curves).

Finally, PIPPS003 was baked at 120°C for 48 hours in vacuum, to remove the high field Q-slope. The result of the test is shown in fig. 19, where for comparison we plot also a standard EP+120°C for 48 hours cavity. The PIPPS003 cavity reached 42 MV/m, limited by quench, with very low level of field emission observed towards the highest fields. This result once again confirms that the baking with caps and foil solution prevents any contamination from depositing on the surface and that by eliminating the material removal step post-furnace there is no negative effect on performance, neither for Q nor for gradients.

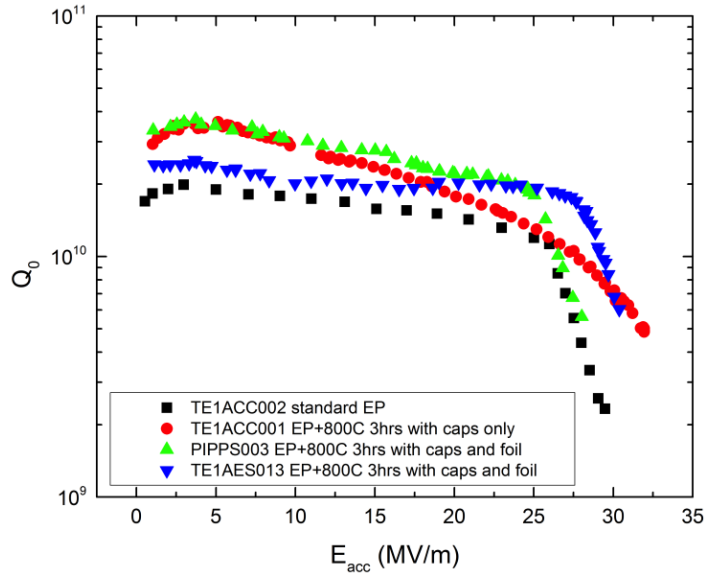

Figure 18. Comparison of post $800^0$C performance for three fine grain cavities baked with the caps solution, versus a standard EP cavity (2K Q-curves).

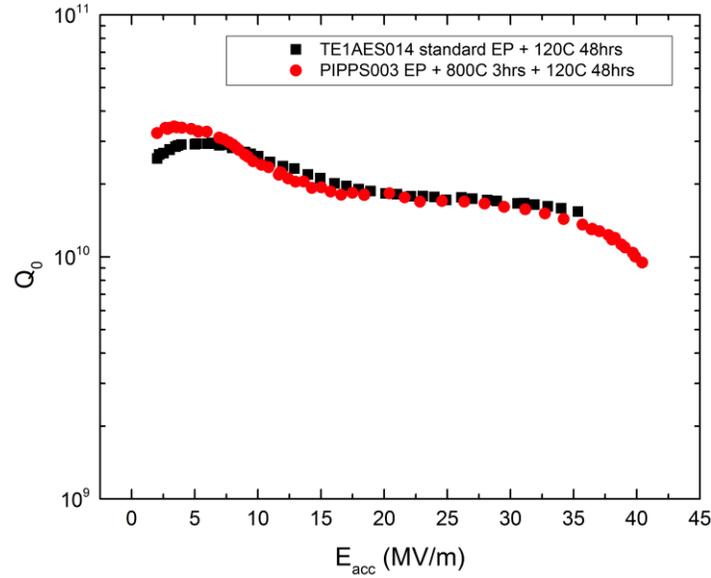

Figure 19. Comparison of Q(2K) curves for an average performing cavity standardly treated with bulk EP plus 120°C for 48 hours, and PIPPS003 baked at 800°C followed by no material removal plus 120°C for 48 hours.

Table 2: Summary of Q(2K) and residual resistance values at 5 MV/m for cavities post 800°C bake for 3 hours. Extremely low values of residual resistances are observed systematically.

| CAVITY ID | Type | Treatment | Q at 5 MV/m, T = 2K | Residual Resistance at 5 MV/m (n$\Omega$) |
|---|---|---|---|---|
| TE1AES016 | Large grain | EP + 800°C 3 hrs no caps, argon venting | 3.5e10 | 1.47±0.44 |
| TE1AES013 | Fine grain | EP + 800°C 3 hrs with caps plus foil, dry air venting | 2.4e10 | <1.09 |
| PIPPS003 | Fine grain | CBP + EP + 800°C 3 hrs with caps plus foil, nitrogen venting | 3.5e10 | 1.45±0.84 |
| TE1ACC001 | Fine grain | EP + 800°C 3 hrs with caps only, nitrogen venting | 3.5e10 | 0.85±0.67 |

*Discussion and outlook*

We have presented a series of cavity experiments and samples investigations aimed at studying the performance of cavities post-degassing treatments and at understanding if the post-furnace material removal step, typically done as a 20-40 microns EP or BCP in standard processing of SRF niobium cavities, could be eliminated, and if that would bring advantages in terms of RF performance. The first tests of cavities that were baked at 800°C for 3 hours in the previously described T-M furnace, showed that performance of fine grain cavities were severely degraded, while two large grain cavities were not affected.

The shape of the Q curves of the fine grain cavities post-furnace treatment resembled strongly those of the known hydrogen Q-disease, and brought us to originally suspect that residual hydrogen was the culprit behind this poor performance. However, a series of tests ruled out hydrogen: a) one of the cavities that had poor performance post-bake, was held at 100 K for 7 hours, but the Q curve did not change compared to that obtained with fast cooldown; b) 5 micron removal via EP post 800°C furnace treatment brought the performance back to standard; c) most importantly, baking with caps - which leaves no line of sight to the internal surface of the cavity- lead to good performance. The bake results with the beamtube caps clearly rule out hydrogen and show that the poor Q derives from contaminants sputtered on the surface. The only atypical contaminant that clearly emerged from the samples analysis via EDX and XPS was titanium. Therefore it is likely that precipitates like α-Ti or some carbide of niobium and titanium, are responsible for the strong field dependent RF losses. The grain size effect observed in cavity performances also seems to match with the contaminated grain boundaries observed in baked samples with laser microscopy. Grain boundaries and lattice irregularities are preferential sites for precipitates to form, and this can explain the difference in performance between large grain and fine grain cavities, since niobium is cold worked in the fine grain case but less in the large grain. Also the fact that only the 'upper' side of the samples shows the discontinuous structures at grain boundaries and intra-grain is in line with the baking with caps results.

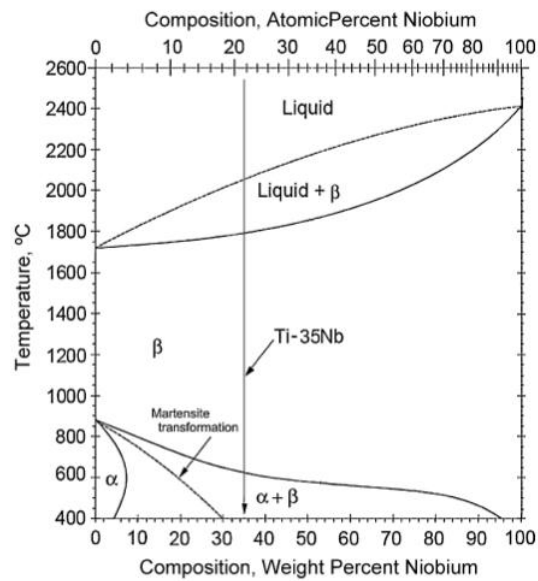

Figure 20. Estimated equilibrium phase diagram of the binary Ti-Nb alloy system [21, 22].

Titanium at grain boundaries was already found to be the culprit of large residual losses and Q-disease-like behaviour by Knobloch as described in [19]. However, in this experimental case titanium was diffusing from the outer cavity wall due to the post-purification process. Even though also in this case hydride precipitation was originally believed to be at fault, evidence of titanium at grain boundaries was found via proton induced x-ray emission by Antoine et al [20]. Knobloch in [19] remarks how, interestingly, this Q-disease-like behaviour is reminiscent of the losses observed in sputtered niobium on copper and NbTiN coated cavities, whose strong medium field losses are believed to be due to the granularity of the films. A model is offered to explain this common behavior, based on intergrain losses, which fits well with the findings and interpretations presented in this paper.

The result on the cavity TE1AES003 which shows improved performance (compared to 800°C for 3 hours) after being baked at 1000°C for 36 hours might still be coherent with the titanium precipitates model. It is possible that a longer baking cycle and at higher temperature, will diffuse titanium deeper in, lowering the concentration at the surface and leading to a solid solution (β phase). This would imply though that the concentration of titanium at the surface is not 'replenished' during the bake cycle. In fig. 20 it is shown a phase diagram for the Ti-Nb alloy [21, 22]. α-Ti is not superconducting at 2K, while β phases are good superconductors. A longer bake at 1000°C might indeed produce a different titanium surface concentration and a different phase formation than a shorter bake at 800°C – leading in the 800°C 3 hours case to both α-Ti and β-Ti formation, and in the 1000°C 36 hours case to only β-Ti formation. One more possibility is that a long 1000°C bake cures much of the niobium cold work induced stress and therefore prevents (or limits) α-Ti precipitates formation.

It is also important to understand the origin of titanium presence in the furnace. Most probably, it comes from the Nb-Ti cavity flanges. At 1000°C the vapour pressure of titanium is ~ $10^{-8}$ Torr, which can be high enough for precipitates to start forming. However at 800°C the vapour pressure of titanium is very low, below $10^{-10}$ Torr. Several cavities were baked at 1000°C in the furnace used for these studies, which makes it seem acceptable that this way the furnace ended up contaminated by titanium. In particular, the line of sight between heating elements and cavity surface should be avoided. At every cycle the titanium might deposit back on the molybdenum strips, as the heat is shut down. At next cycle, the heating elements are the first to warm up, and therefore the titanium may leave the molybdenum surface and sputter on any line of sight surface.

The beamtube caps together with the additional wrap-around niobium foil lead to surface resistances even better (lower) than the ones produced after standard EP. These results prove therefore that the post high temperature baking losses problem is not due to any 'intrinsic' limitation (like for example oxygen diffusion) but rather to extrinsic factors: it arises from furnace contamination, and eliminating any line of sight to the internal cavity surface can circumvent the problem. We also are able to pinpoint the source of the contamination in our furnace, which is titanium, likely coming from the Nb-Ti flanges. Even though different furnaces for SRF cavities degassing might have contaminants of different nature and in different amount, the titanium is a likely common issue due to the fact that most of SRF cavities are fabricated with Nb-Ti flanges. These results therefore suggest that precautions should be taken when baking cavities with such flanges to avoid contaminating the furnace. Moreover, the beamtube caps solution is functional for any form of contaminant and can therefore be implemented as an inexpensive routine tool for the high temperature treatment of SRF cavities, making the post-degassing material removal unnecessary.

Finally, an extremely important result emerges from these preliminary studies: fine grain cavities which have been baked at 800°C for 3 hours with caps (and the large grain with no need of caps), followed by no chemistry, consistently show extraordinary low values of the low field residual resistances ~ 1 nΩ and below. The low field BCS resistance is in most cases (TE1AES016, TE1ACC001, PIPPS003) ~ 6-7 nΩ at 2 K, close to the minimum value typically reached via 120°C bake. A low value of BCS resistance combined with an extremely low value of residual resistance make an excellent low field $Q(2K) \sim 3.5\times10^{10}$ emerge for all the three cavities. For the cavity TE1AES013 instead the value of BCS resistance is the standard one for an EP cavity, which is ~ 10 nΩ. This could originate from the fact that the cavity, differently from the other ones, was the only one vented with dry air at room temperature. We plan on further studies with a larger set of cavities and for different baking temperature range in the future. Field dependences of the components of the surface resistances [23] for these cavities are currently under study and will be subject of future publications.

The successful elimination of the final material removal step post degassing treatment in a vacuum furnace with a clever, simple and inexpensive solution like the beamtube caps, not only provides a cost effective alternative to the current established cavity processing stream, but also brings large advantage for the RF performance, of particular importance for cavities for CW SCRF machines.

*Conclusions*
We have studied the origin of the RF losses in niobium cavities post degassing in a high temperature furnace. Results indicate that furnace contaminants like titanium cause formation of lossy precipitates in stress/grain boundaries rich cavities. Baking the cavities with the presented solution of beamtube caps, which prevent contaminant from reaching the Nb surface, offer a solution to circumvent the problem. These findings help reaching a cost effective alternative by cutting some processing steps; they also open the possibility for environmentally friendly cavity acid free processing via centrifugal barrel polishing, and lead to an improvement of the RF performance by lowering systematically the cavity residual resistance to atypically low values of ~ 1 nΩ and below.

*Acknowledgements*
Authors would like to acknowledge Hasan Padamsee for suggesting the experiment of baking the cavities wrapping the flanges with niobium foil. We also thank Slava Yakovlev for initiating and supporting the high Q R&D, and the T&I department for the support with cavity testing. The work was partially supported by the DOE Office of Nuclear Physics. Fermilab is operated by Fermi Research Alliance, LLC under Contract No. DE-AC02-07CH11359 with the United States Department of Energy.